\documentclass[a4paper,11pt]{article}
\usepackage{pos}
\usepackage{wrapfig}
\usepackage{physics}
\usepackage{lineno}
\usepackage{siunitx}
\DeclareSIUnit \solarmass {M_{\odot}}
\DeclareSIUnit \parsec {pc}

\usepackage{csquotes}
\usepackage{multirow}

\begin{document}

\title{Prospects for the Detection of the Standing Accretion Shock Instability in IceCube-Gen2 
}
\ShortTitle{SASI Prospects in IceCube-Gen2}

\author{Jakob Beise, for the IceCube-Gen2 Collaboration} 

\emailAdd{jakob.beise@icecube.wisc.edu}

\abstract{Core-collapse supernovae (CCSNe) are among the most energetic processes in our Universe and are crucial for the understanding of the formation and chemical composition of the Universe. The precise measurement of the neutrino light curve from CCSNe is crucial to understanding the hydrodynamics and fundamental processes that drive CCSNe. The IceCube Neutrino Observatory has mass-independent sensitivity within the Milky Way and some sensitivity to the higher mass CCSNe in the Large and Small Magellanic clouds. The envisaged large-scale extension of the IceCube detector, IceCube-Gen2, opens the possibility for new sensor design and trigger concepts that could increase the number of neutrinos detected from a CCSNe burst compared to IceCube. In this contribution, we study how wavelength-shifting technology can be used in IceCube-Gen2 to measure the fast modulations of the neutrino signal due to standing accretion shock instabilities (SASI).

\vspace{4mm}
{\bfseries Corresponding author:}
Jakob Beise$^{1}*$\\
{$^{1}$ \itshape 
Department of Physics and Astronomy, Uppsala University, Box 516, S-75120 Uppsala, Sweden
}\\
$^*$ Presenter

\ConferenceLogo{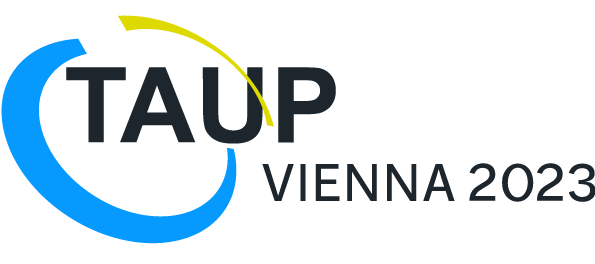}

XVIII International Conference on Topics in Astroparticle and Underground Physics 2023

\FullConference{%
XVIII International Conference on Topics in Astroparticle and Underground Physics 2023 (TAUP 2023)\\
  28 August -- 1 September, 2023\\
  Vienna, Austria}
}

\maketitle

\section{Introduction}
\label{sec:int}

\noindent The IceCube Neutrino Observatory \cite{IceCube:2016inst} is a Cherenkov detector instrumenting \SI{1}{\kilo \m} of glacial ice at the geographical South Pole. Consisting of 5160 optical sensors on 86 vertical cables, called \enquote{strings}, buried \SI{1450}{\m} to \SI{2450}{\m} underneath the ice, IceCube has detected neutrinos ranging in energy from \SI{}{\giga \eV} to \SI{}{\peta \eV}. Though IceCube's primary strength is the study of high-energy astrophysical neutrinos and their sources, IceCube has sensitivity to \SI{}{\mega \eV} supernova (SN) neutrinos on a statistical basis.

IceCube is primarily sensitive to the $\overline{\nu}_e$ flux which produces positrons via inverse-beta decay. Due to the sparse detector instrumentation, Cherenkov photons generated by individual secondary positrons will be picked up by at most one Digital Optical Module (DOM) in the vast majority of cases with a sensor dark noise rate of about \SI{540}{\Hz} per DOM. Instead, a galactic CCSNe can be observed as an excess of the collective detector rate over the background for a few $\mathcal{\SI{10}{\s}}$ during the accretion and cooling phase of a CCSNe \cite{IceCube:2011pros}. Ultimately, for SN detection in IceCube, the signal-to-noise ratio is paramount. IceCube can neither identify the neutrino flavour nor reconstruct the direction in the low-energy regime. On the other hand, due to the high statistics of the detected signal, IceCube has excellent resolution for detecting fast-time features in the supernova light curve \cite{IceCube:2011pros}. IceCube has sensitivity above $10\sigma$ independent of the progenitor mass within the entire Milky Way and limited sensitivity in the Large and Small Magellanic Clouds (LMC and SMC) \cite{IceCube:2011pros}.

IceCube-Gen2 \cite{IceCube-Gen2:2023tdr} is an envisioned large-scale extension adding a total of 9,600 new optical sensors on 120 new vertical \enquote{strings} with 80 modules each, spanning depths between \SI{1369}{\m} and \SI{2689}{\m}. With an inter-module vertical spacing of \SI{17}{\m} and horizontal spacing of \SI{240}{\m} the detector is too sparse to observe a single inverse-beta decay in more than one module. IceCube-Gen2 will utilize segmented sensors housing multiple, small-diameter Photomultiplier Tubes (PMT) which, given the increased photocathode area, will enhance photon collection as well as sensor noise suppression. This will extend the sensitivity to low mass CCSNe out to the LMC and SMC \cite{LozanoMariscal:2021mult, IceCube-Gen2:2023tdr}.  

Wavelength shifters (WLS) shift the UV contribution of the Cherenkov spectrum into the visible regime, thereby augmenting the photon collection of a module. Because of the relatively small read-out PMTs little sensor noise is added, which further improves the signal-to-noise ratio in SN detection. A Wavelength-Shifting Optical Module (WOM) \cite{WOM:2022ini, WOM:2022upgr} is currently developed and a total of 12 modules will be deployed in the IceCube Upgrade \cite{Ishihara:2019}, a planned dense infill of roughly 700 new sensors in the DeepCore region targeted to improve detector calibration and obtaining world leading measurements on neutrino oscillations. For IceCube-Gen2, WLS tubes as an add-on to the already planned segmented sensors have been discussed as a cost-efficient photon collector. Such a passive component would operate without electronics or PMT readout and instead be viewed by PMTs on another module. Because of the absence of a pressure housing, these tubes would be more efficient and larger geometries could be achievable. In addition, the only noise contribution of the WLS tubes would come from radioactive decays in the tube material.

In Ref.~\cite{WOM:2023appl} it was noted, that WLS can improve the sensitivity towards detecting CCSNe. In this contribution, we study the capability to distinguish a SN light curve featuring the fast, time-varying standing accretion shock instability (SASI), arising from hydrodynamical instabilities during the CCSN accretion phase, from a \enquote{flat} light curve exhibiting no modulations. Section~\ref{sec:ana} describes the analysis method, Sec.~\ref{sec:res} presents the results and in Sec.~\ref{sec:con} we summarise our findings.

\section{Analysis}
\label{sec:ana}

\noindent For this study we use \texttt{SNEWPY} \cite{SNEWS:2021} to simulate the initial neutrino flux of the Tamborra 2014 \SI{20}{\solarmass} model \cite{Tamborra:2014sasi} during the entire simulation window from \SI{6}{\milli \s} to \SI{338}{\milli \s} post-bounce. The default case in this study assumes the optimistic case of no flavour mixing. We use \texttt{ASTERIA} \cite{ASTERIA} to simulate the detector response for three detector geometries which we will refer to as IceCube (incl. DeepCore), Gen2 (IceCube-Gen2 excluding WLS) and Gen2+WLS (IceCube-Gen2 including WLS) in the following. The baseline design for IceCube-Gen2 uses Long Optical Modules (LOMs) \cite{Basu:2021lom, IceCube-Gen2:2023tdr}, multi-PMT modules that are optimised for low-power consumption and that fit narrower bore holes. However, since the properties of the LOM are still being characterised we use the multi-PMT Digital Optical Module (mDOM) \cite{Kossatz:2017mdom, IceCube-Gen2:2023tdr} in this study. The WLS component of this study considers a \SI{2}{\m} long tube with an outer diameter of \SI{256}{\milli \m}, a tube thickness of \SI{10}{\milli \m} and the same material properties as the inner tube of the WOM \cite{WOM:2022ini}.

The time-dependent number of detected photons from a CCSN in all detector components can be expressed as the double sum over all neutrino and antineutrinos of all flavours $l=e, \mu, \tau$ and detector components $i$ of the product of the time-dependent neutrino density $n_{\nu_l}(t)$, the number of modules $m_i$, the energy-dependent number of radiated Cherenkov photons $N_{\gamma, l}(E_l)$ by a lepton of energy $E_l$ and the depths-averaged, single photon effective volume $\overline{V_{\gamma, i}^{\text{eff}}}$ \cite{IceCube:2011pros}:

\begin{equation}
    N_{\text{SN}}(t) = \sum_l \sum_i n_{\nu_l}  \cdot m_i \cdot \epsilon_{\tau, i} \cdot N_{\gamma, l} \cdot \overline{V_{\gamma, i}^{\text{eff}}} \ ,
\end{equation}

\noindent where $\epsilon_{\tau, i}(R_{\text{SN}}) = \epsilon_{\tau, i}^{\text{max}}/(1+R_{\text{SN}} \tau)$ is the rate-dependent dead-time efficiency for a non-paralysing dead time $\tau$. The number of background hits during a time window $\Delta t$ is the sum over all noise rates $R_{\tau, i}$:


\begin{equation}
    N_{\text{bkg}} = \sum_i m_i \cdot \Delta t \cdot R_{\tau, i} \ .
\end{equation}

\begin{table}[htbp]
    \begin{center}
    \caption{Characteristics of the modules considered in this study. $\epsilon_{\tau}^{\text{max}}$ is simulated for a \SI{250}{\micro \s} dead time.}
    \label{tab:sen}
    \begin{tabular}{ccccc}
    module & $m$ & $\overline{V_{\gamma}^{\text{eff}}}$ [\SI{}{\m^{3}}] & $R_{\tau}$ [Hz] & $\epsilon_{\tau}^{\text{max}}$ [\%]\\
    \hline
    DOM & 4800 &  0.17 & 285 & 88.3\\
    HQE DOM & 360 & 0.23 & 359 & 84.6\\
    mDOM & 9760 & 0.33 & 2300 & 95.8\\
    mDOM + 2 WLS & \ \ 9760 & 0.60 & 2700 & 95.8\\
    \end{tabular} \\
    \end{center}
\end{table}

Table~\ref{tab:sen} lists all relevant data for the modules considered in this study. In the case of Gen2+WLS we add two WLS tubes: one above and one below the mDOM. This nearly doubles $\overline{V_{\gamma}^{\text{eff}}}$, while the noise rate only increases by \SI{400}{\Hz}, or 17\%. For the time resolution we chose $\Delta t = \SI{1}{\milli \s}$.

\begin{figure}[htbp]
    \centering
    \includegraphics[width=0.85\textwidth]{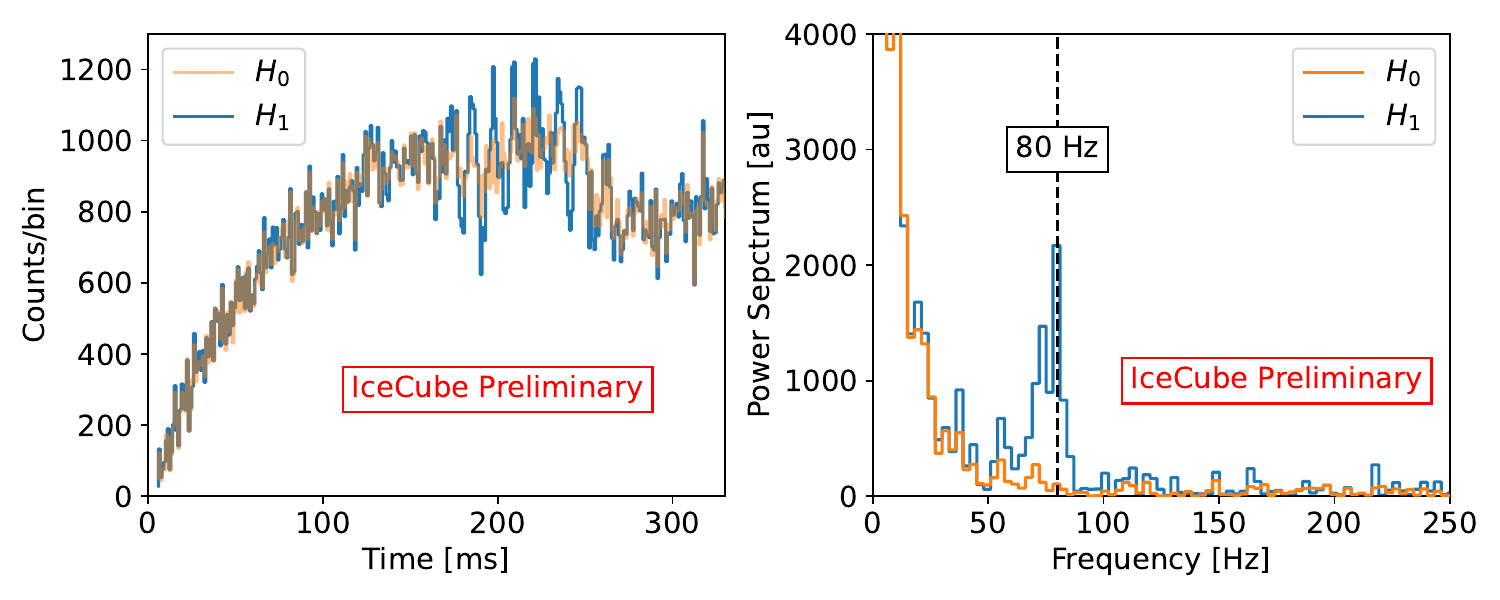}
    \caption{(Left): Number of hits in IceCube for the Tamborra 2014 \SI{20}{\solarmass} model \cite{Tamborra:2014sasi} at \SI{10}{\kilo \parsec}. (Right): Power density spectrum for the same model. The null hypothesis $H_0$ is shown in blue and the signal hypothesis $H_1$ in orange.}
    \label{fig:hypothesis}
\end{figure}

We denote $H_0$, the null hypothesis, as the case in which a CCSNe is detected but whose light curve \textit{does not} feature SASI modulations, and $H_1$, the signal hypothesis, as the case in which a CCSNe is detected and whose light curve \textit{does} feature SASI modulations. Fig.~\ref{fig:hypothesis} (left) shows the number of detected photons in IceCube for the Tamborra 2014 \SI{20}{\solarmass} model \cite{Tamborra:2014sasi} at \SI{10}{\kilo \parsec} corrected by the average background rate. We take the discrete Fourier transform of the \SI{1}{\milli \s}-binned light curves of $H_0$ and $H_1$ and compute the power spectrum (see Fig.~\ref{fig:hypothesis} right). To construct the test statistic (TS) distribution we select the maximum in the power spectrum for frequencies larger than \SI{75}{\Hz} for 10,000 random noise realisations. This cutoff frequency is motivated by the \SI{80}{\Hz} SASI frequency predicted in Ref.~\cite{Tamborra:2014sasi}.

\section{Results}
\label{sec:res}

\begin{figure}[htbp]
    \centering
    \includegraphics[width=\textwidth]{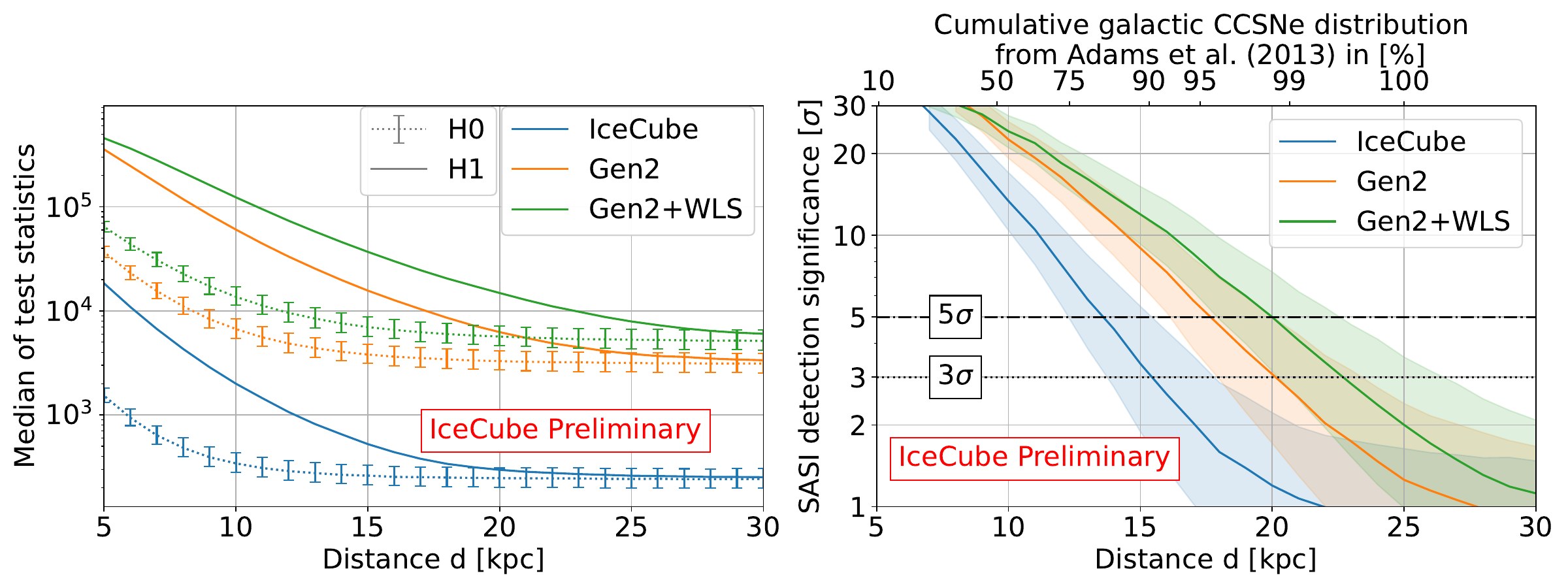}
    \caption{(Left) Median and 16\%-84\% quantiles of the $H_0/H_1$ TS distribution as a function of progenitor distance. (Right) SASI detection significance over distance shown for IceCube, Gen2 and Gen2+WLS.}
    \label{fig:result}
\end{figure}

\noindent When we repeat the procedure over a range of source distances for $H_0$ and $H_1$ we obtain Fig.~\ref{fig:result} (left), which displays the median (line) and 16\% to 84\% quantiles (size of error bars) of the TS distribution as a function of distance. We note that the error bars for the signal hypothesis are not displayed to improve visibility. We can directly infer the p-value and two-sided significance of the deviation which is plotted in Fig.~\ref{fig:result} (right), where the coloured bands indicate the 16\% and 84\% quantiles. We also show the cumulative galactic CCSNe distribution from Ref.~\cite{Adams:2013ccsn} to translate the increase in reach into coverage. As can be seen, IceCube-Gen2 with WLS will cover more than 98.5\% of the Milky Way at $5\sigma$ compared to 82.8\% in IceCube alone.

We note that the results obtained are sensitive to the analysis method and analysis cuts set, the assumed neutrino flavour mixing and the CCSNe model. Regarding the analysis cuts, we ran the same analysis with tighter frequency cuts ($\SI{75}{\Hz} < f < \SI{85}{\Hz}$), time windows centred around the SASI period ($\SI{150}{\milli \s} < f < \SI{300}{\milli \s}$) and a combination of both. For the flavour mixing, we considered the most pessimistic case of complete flavour exchange as well as the MSW effect for normal and inverted hierarchy. Finally, we also considered the Tamborra 2014 \SI{27}{\solarmass} model \cite{Tamborra:2014sasi} for three different observer directions $d_1, d_2, d_3$ relative to the SASI plane.

\begin{figure}[htbp]
    \centering
    \includegraphics[width=0.7\textwidth]{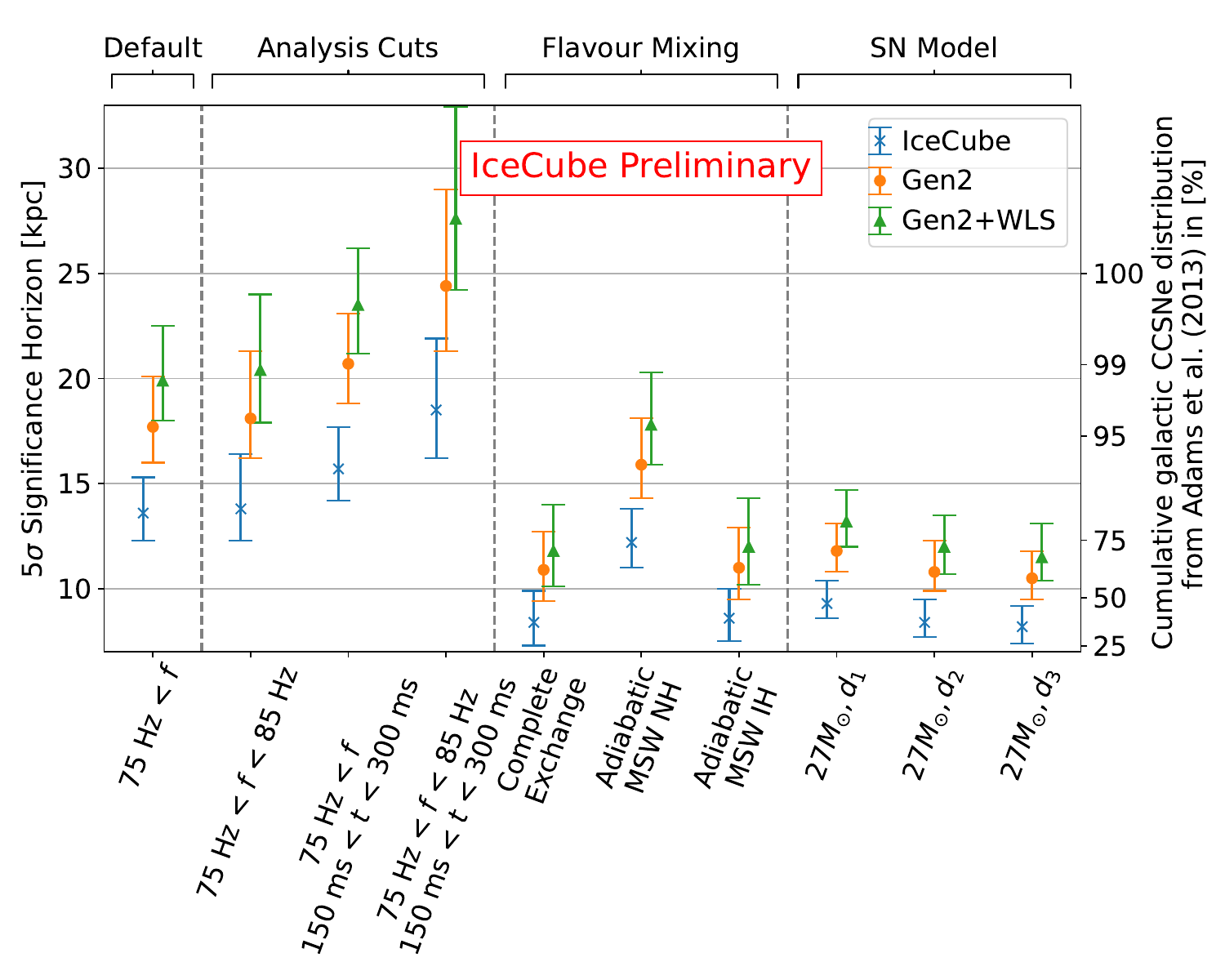}
    \caption{Compilation of all systematic modification to the default setting.}
    \label{fig:systematics}
\end{figure}

Figure~\ref{fig:systematics} shows the $5\sigma$ significance horizon as well as the galactic CCSNe coverage for all considered simulation setups. The results of Fig.~\ref{fig:result} are shown as \enquote{Default}, while the adaptation from that setting is indicated on the x-axis. We find that the stronger the analysis cut, the larger the reach but the lesser the capability to generalise the study to different models. We also find that a more realistic treatment of flavour conversion results in decreased coverage. Finally, the \SI{27}{\solarmass} model yields a much-reduced significance horizon. 

\section{Conclusion}
\label{sec:con}

\noindent IceCube has a unique capability to observe small-scale modulations of the CCSN light curve with high precision giving valuable insights into the hydrodynamics of exploding stars. In this contribution, we demonstrated the potential of IceCube-Gen2 and the use of wavelength shifters to detect the effect of SASI on the SN light curve.
While the improvement of adding WLS is marginal for the detection of SASI, which is already a significantly strong signal in IceCube-Gen2, there are potential opportunities to enhance fainter signals of fast-time variations (i.e. from rotational CCSNe). The procedure outlined here can also be further used for studies of generic models of SASI modulation such as a neutrino light curve with a custom frequency and amplitude.

\bibliographystyle{ICRC}
\bibliography{references}

%
%
%


\end{document}